\documentclass[aps,prl,reprint,superscriptaddress]{revtex4-1}
\usepackage{blindtext}
\usepackage{graphicx}
\usepackage{amssymb,amsfonts,amsmath}
\usepackage{epstopdf}
\usepackage[utf8]{inputenc}

\begin{document}
\author{Diego Baresch}
\affiliation{Sorbonne Universités, UPMC Univ Paris 06, CNRS, UMR 7588, Institut des Nanosciences de Paris, F-75005, Paris, France.}

\author{Jean-Louis Thomas}
\email{jean-louis.thomas@upmc.fr}
\affiliation{CNRS, UMR 7588, Institut des NanoSciences de Paris, F-75005, Paris, France}

\author{Régis Marchiano}
\affiliation{Sorbonne Universités, UPMC Univ Paris 06, CNRS, UMR 7190, Institut Jean le Rond d'Alembert, F-75005, Paris, France.}

\title{Observation of a single-beam gradient force acoustical trap for elastic particles: acoustical tweezers}



\begin{abstract}
We demonstrate the trapping of elastic particles by the large gradient force of a single acoustical beam in three dimensions. Acoustical tweezers can push, pull and accurately control both  the position and the forces exerted on a unique particle. Forces in excess of 1 micronewton were exerted on polystyrene beads in the sub-millimeter range. A beam intensity less than 50 Watts/cm$^2$ was required ensuring damage-free trapping conditions. The large spectrum of frequencies covered by coherent ultrasonic sources provide a wide variety of manipulation possibilities from macro- to microscopic length scales. Our observations could open the way to important applications, in particular in biology and biophysics at the cellular scale and for the design of acoustical machines in microfluidic environments. 
\end{abstract}

\maketitle

Arthur Ashkin observed that the radiation pressure of a focused laser could significantly accelerate particles in the direction of light propagation \cite{Ashkin70}. An acceleration that results from momentum conservation and light reflection by the particle. Importantly, an additional unanticipated force transversally attracted the particles towards the beam axis where they were successively optically guided. This component of radiation pressure was called gradient force. 
Axial pushing forces could be canceled by the wall of the chamber or by a second counter-propagating beam in order to obtain a stable equilibrium position. Only two decades later did the first single-beam gradient trap appear, a specific scheme dubbed optical tweezers (OTs) \cite{Ashkin2011},\cite{Ashkin86}. OTs can attract and stably trap a particle at a single equilibrium position in space, the beam focus. Defying intuition, the light beam exerts a negative pulling force on a particle located downstream from the focus. This restoring force is the axial counterpart of the radial gradient force initially observed. The particles can be displaced with nanometer accuracy by simply moving the light beam mechanically. These unique features of OTs lead to dexterous contactless manipulation in three dimensions and have had a wide and far-reaching impact since their inception, spanning the range from life science, material science and microfluidics to laser cooling of atoms \cite{Ashkin87Nature,Grier,Dholakia,Padgett2011,Tannoudji}. Moreover, precisely calibrated OTs have enabled force measurements with piconewton accuracy on single molecules and opened up important new areas in biophysics \cite{Svoboda1994}.

An early recognized issue is the heating of both the object and the trapping medium \cite{Neuman}. Radiation pressure is proportional to the wavefield intensity divided by its speed of propagation - $I/c$. Hence, light intensities of about $10^7-10^8$~W/cm$^2$ at the focus yield weak forces in the pN range on micron sized objects \cite{Ashkin1980}. From this point of view, introducing ultrasonic traps is relevant. The change in nature of the oscillating field widens the prospect of operating in media generally exposed to heating by light. The speed of sound gives a major advantage to acoustics (5 orders of magnitude). Furthermore, piezoelectric sources with high efficiency are available with frequencies from kHz up to GHz, scaling macroscopic to microscopic wavelengths, broadening the range of particle sizes that can be trapped as well as the operating distances from the source. 

Acoustic levitation traps were developed independently and concomitantly with their optical counterpart \cite{Eller68,Trinh85,Apfel81}. Recent improvements to standing acoustic wave schemes lead to levitating and translating single or multiple particles in air \cite{Foresti} and acoustophoresis provides advanced particle, cell and organism separation in complex microfluidic environments \cite{Laurell,Bruus}. Standing wave schemes have recently been proposed to accurately manipulate particles in two dimensions using surface \cite{Ding,Tran} or bulk acoustic waves \cite{Demore2} with phase or frequency shifts in order to demonstrate capabilities similar to OTs. 
However, all the aforementioned techniques share the same limitations, e.g. standing waves form multiple equilibrium positions in one or two dimensions each of which is likely to trap one or various particles at the same time, therefore precluding  separability and selectivity at the single particle level with ease \cite{Huang2,Demore3}.  

Obtaining a three dimensional restoring force by tightly focusing an acoustic beam has proven to be challenging: Wu reported on a dual beam trap \cite{Wu} whereas lateral manipulation with an ultrasonic beam was successfully implemented while the axial expelling force was canceled by a membrane \cite{Lee}. Both concepts are similar to those reported in Ashkin's seminal work on particle acceleration \cite{Ashkin70}.

Here we report on the first observation of a negative gradient pulling force with acoustic waves and hence the achievement of all acoustical single beam trapping ; i.e. an acoustical tweezer.

From our previous work, we inferred that failure to trap in previous attempts were expected with an axisymmetric beams. Indeed, solid elastic particles have a propensity to be expelled from intensity maxima at the focal spot of a conventional focused ultrasonic beam. Starting from recent developments of the theory for acoustic radiation force (ARF) calculations \cite{Marston2006,Marston2009} and the acoustical analogous to the generalized Lorenz-Mie theory, \textit{GLMT} \cite{BareschJASA,Gouesbet}, it can be shown (see supplementary text \cite{Supp}) that for small spherical particles (radius $a$) compared to the wavelength ($\lambda$), the ARF takes the following form:
\begin{eqnarray}
\vec{F}&=& -\frac{1}{2}\left\{\Re(\alpha_m)\frac{1}{2\rho c^2}\vec{\nabla}|p|^2 - \Re(\alpha_d)\frac{1}{2}\rho\vec{\nabla}|\vec{v}|^2 \right.  \nonumber \\
&&\left. + \left( \frac{k}{c}\Im(\alpha_m)-\frac{k^4}{12\pi c}\Re(\alpha_m)\Re(\alpha_d) \right)\Re(p\vec{v}^\ast) \right. \nonumber \\
&&\left. + \rho\Im(\alpha_d)\Im((\vec{v}\cdot\vec{\nabla})\vec{v}^\ast)\right\} \label{eq:force}
\end{eqnarray}
where $p$ and $\vec{v}$ are the first order acoustic pressure and particular velocity fields respectively ; $\rho$ is the host fluid's density and $k=2\pi f/c$ ($f$ is the frequency and $c$ is the wave speed). $\Re$ and $\Im$ denote the real and imaginary parts of these complex fields while $^*$ stands for complex conjugation.
$\alpha_m= \alpha_m^0/(1+i\frac{k^3}{4\pi}\alpha_m^0)$ and $\alpha_d= \alpha_d^0/(1-i\frac{k^3}{12\pi}\alpha_d^0$) are two acoustic strength parameters associated to the sphere's monopolar and dipolar modes where:
\begin{eqnarray}
\alpha_m^{0}&=&\frac{4}{3}\pi a^3\left(1-\frac{\rho c^2}{\rho_p (c_l^2-\frac{4}{3}c_t^2)}\right)\\
\alpha_d^0&=&4\pi a^3\left(\frac{\rho_p -\rho}{2\rho_p + \rho}\right)
\end{eqnarray}
$c_l$ and $c_t$ are the longitudinal and transverse propagation speeds in the particle of density $\rho_p$. The first two terms in Eq.(\ref{eq:force}) stand for an acoustic gradient force that is tantamount to its optical counterpart while the remaining terms form the scattering force. For most of elastic particles suspended in liquids, the gradient term associated to an axisymmetric beam points towards energy density minima precluding from establishing a trapping behavior. However, we have recently proposed that shaping the beam's wavefront may restore three dimensional trapping forces \cite{BareschJAP} as a consequence of a monopolar mode annihilation on the propagation axis (see SI text \cite{Supp}).
Focused acoustical vortices satisfy various constraints to retrieve both radial and axial restoring gradient forces and such peculiar wave fields have been successfully generated in acoustics \cite{Hefner,MarchianoThomas2003,MarchianoThomas2008} and applied to investigate the transfer of orbital angular momentum to matter \cite{Volke,PadgettAcous,Demore,Wunenburger,Demore2}. 

\begin{figure}
\centering \includegraphics[width=1.0\linewidth]{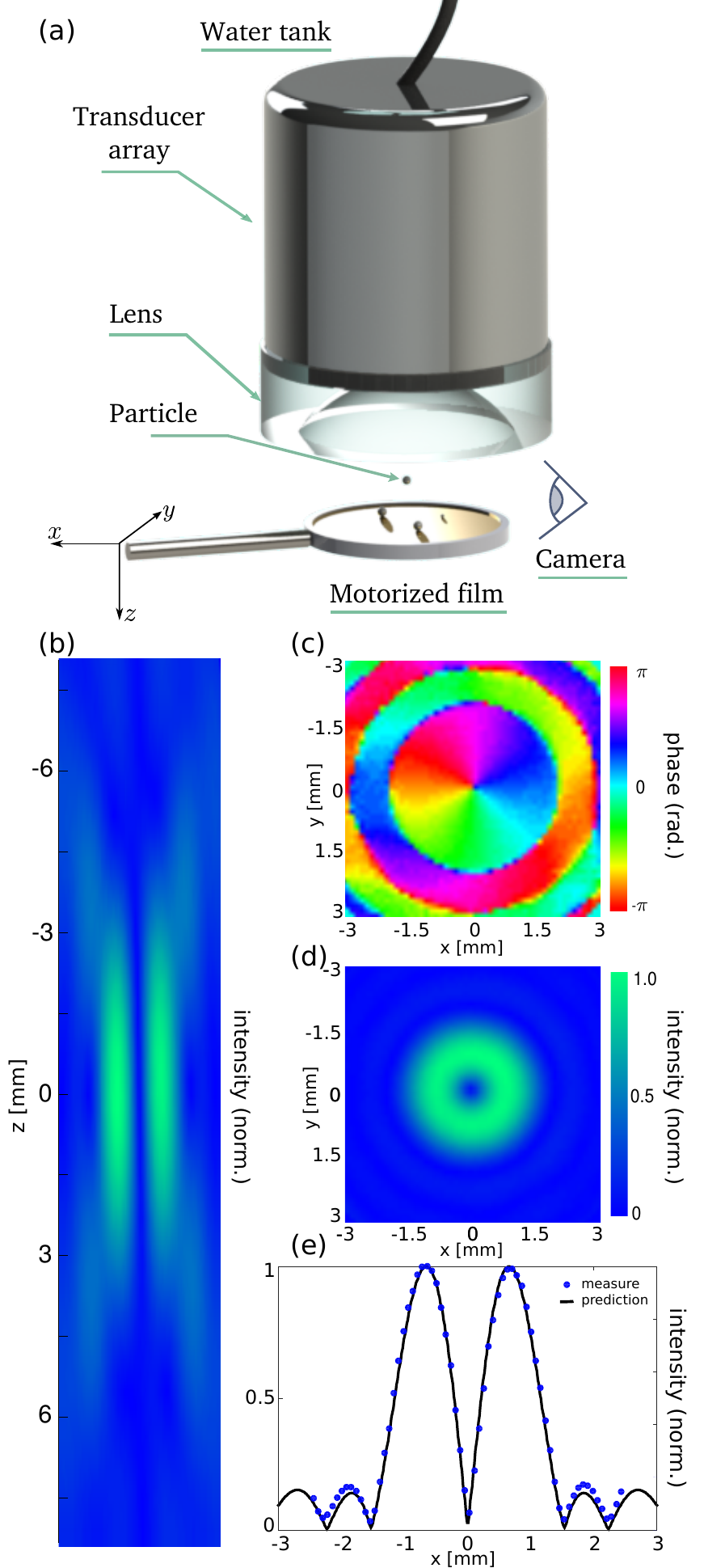}
\caption{(Color online) (\textbf{a}), Sketch of the ultrasonic emitter (1.15 MHz) immersed in a water tank. The beam is focused with a concave acoustical lens of \textit{f}-number $\simeq0.7$ upon polystyrene particles lying on an acoustically transparent polyethylene film. (\textbf{b}), Colorplot of the normalized acoustic intensity in the direction of propagation (negative to positive $z$). (\textbf{c-d}), Field's phase and intensity  in the focal plane. The variation of the phase from $-\pi$ to $\pi$ around the propagation axis indicates structured helical wavefronts. The ring pattern is a peculiarity of vortex beams. (\textbf{e}), Comparison of the measured and predicted pressure fields (intensity) for $y=0$~mm.}
\label{fig:vors}
\end{figure}

In Fig.~\ref{fig:vors} a), a sketch of the experimental setup is shown. A multi-element array of piezoelectric transducers is driven by independent arbitrary signal generators to synthesize the beam in a large water tank. A tight focusing of the vortex is obtained with a high \textit{f}-number acoustical lens. Based on numerical calculations, \cite{BareschJAP}, an optimal acoustical vortex is determined and used as an input to a versatile and robust wavefront synthesis technique \cite{MarchianoThomas2003}. The array of transducers is driven with the calculated signals to experimentally generate the incident field. We drive the transducers at a frequency $f=1.15$~MHz corresponding to a wavelength $\lambda=1.3$~mm in water. 
A hydrophone scanning a grid of points defined near the focus measures the instantaneous pressure field. The normalized fields intensity in the direction of propagation is shown in Fig.~\ref{fig:vors} b). The depth of field is approximatively equal to $4\lambda=5.2$~mm and will determine the axial trapping extent of the tweezers. In the focal plane (Fig.~\ref{fig:vors}c-e), the acoustical intensity is focused to a ``ring" pattern that is a peculiarity of vortex beams. Looking at the phase diagram (Fig.~\ref{fig:vors}c), its variation from $-\pi$ to $\pi$ around the beam axis reveals the helicoidal structure of the wavefronts. Since a single jump of $2\pi$ is achieved, the pitch of the helix is equal to $\lambda$ and is here left-handed. Measured and predicted pressure fields are in perfect agreement (Fig.~\ref{fig:vors}e) and Figs.~S1, S2 \cite{Supp}). 
When the emitter is driven at  maximum power, the peak value of the pressure on the ring reaches $0.8\pm$0.1~MPa and the acoustical intensity of approximatively $42$~W/cm$^2$ remains weak at the focus.

\begin{figure}
\centering 
\includegraphics[width= 0.65 \linewidth]{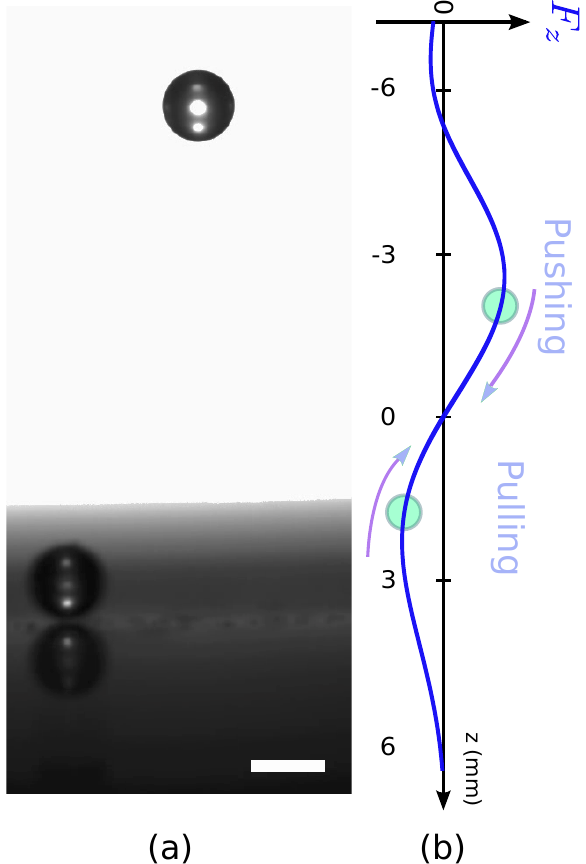}
\caption{(Color online) (\textbf{a}), Photograph of a particle of radius $a=170\,\mu$m trapped in three dimensions. The large negative axial radiation force lifts the particle from a distance $\simeq2.6$~mm below the equilibrium position. A second particle lying on the film is not attracted towards the focus. The scale bar represents $200\,\mu\mbox{m}$. (\textbf{b}), Theoretical prediction of the axial acoustical radiation force as a function of the particle's position on the propagation axis.  }
\label{fig:vertical}
\end{figure}

\paragraph{Acoustical tweezers}
In this vertical configuration, the ultrasonic emitter focuses the beam upon a thin polyethylene film (largely transparent to acoustic waves) on which the polystyrene particles are dispersed (their diameters range from $190\mu\mbox{m}$ to $390\mu\mbox{m}$). The film, mounted on a three-axis positioning system, allows to move the particles around the region of interest. As the downward propagating beam impinges a particle, the latter is lifted by a strong negative gradient force and remains stably trapped in three dimensions. A photograph of a levitated particle (radius $a=170\mu$m) is shown in Fig.~\ref{fig:vertical} a) (see SI Video 1 \cite{Supp} as well). The predicted behavior of the axial force is shown in Fig.~\ref{fig:vertical} b). It is negative beneath the beam focus (lifting force) and positive above (pushing force). Nevertheless, the particle reaches an axial equilibrium position when the negative radiation force balances adverse effects acting in the direction of propagation and the axial equilibrium position is generally below the beam's focus (typically $30.0$~mm away from the lens). These opposite forces are the positive axial scattering force, the gravity and the viscous drag force resulting from the acoustic streaming, that is to say a flow generated by acoustic absorption of the beam's intensity in the bulk of the fluid \cite{Eckart,Riaud}. 
The only weight of the sphere represents $20$~nN for the largest polystyrene particles. The axial range of operation of the tweezers is investigated by changing the distance separating the particle and the beam focus before switching the source on. The maximum lifting distance was found to reach $3.4$~mm or equivalently $\simeq 2.6\lambda$ beneath the focus for the maximum emission intensity ($I=42$W/cm$^2$). Contrarily, if the particle is initially located above the equilibrium position, it is pressed against the film by a positive axial force. 

\begin{figure}
\centering \includegraphics[width=0.85\linewidth]{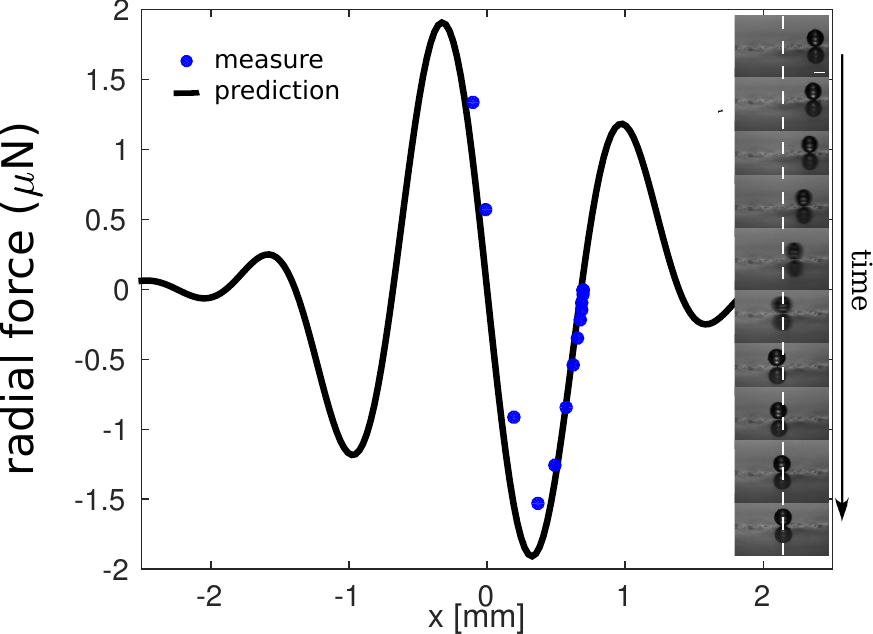}
\caption{(Color online). Measurement of the radial force as a particle is accelerated towards the propagation axis. The force is obtained as a function of the radial position $x$ by calculating the radial acceleration from the instantaneous position of the center of the sphere.}
\label{fig:rad}
\end{figure}

\begin{figure}
\centering \includegraphics[width=0.85\linewidth]{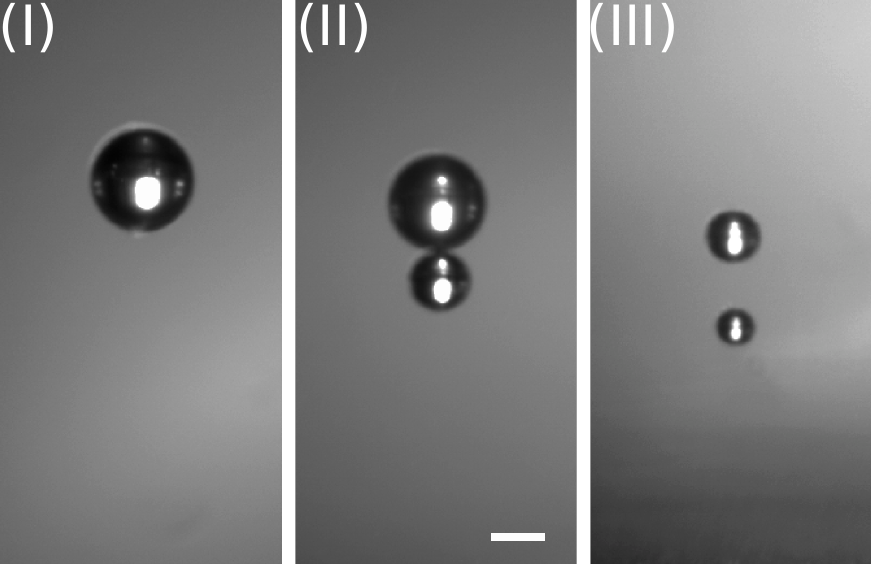}
\caption{Systems of one or two particles can be trapped. (I) radius $a=189~\mu\mbox{m}$. (II), radii $a=177$ and $108~\mu\mbox{m}$.  (III), radii $a=98$ and $67~\mu\mbox{m}$.  The scale bar represents $200\,\mu\mbox{m}$. Their self-arrangement exhibits the effect of secondary acoustic radiation forces.}
\label{fig:multi}
\end{figure}

The radial component of the radiation force is predicted to be at least one order of magnitude stronger than the axial counterpart (See Fig.S3 \cite{Supp}). So, when the particle is initially located aside the beam axis, a large radial acceleration is observed before it starts to lift. Acquiring images at a high frame rate (520 fps), we recorded the motion of the particle as it is attracted towards the central beam core.  
By taking into account the viscous drag force and other inertial forces arising from the particle's acceleration (see Fig.~\ref{fig:rad} and Supplementary text), we calculated the net radial  force acting on the particle. For a polystyrene particle of radius $a=190\,\mu$m the radial force reaches $F_\rho\simeq1.5\,\mu$N at a radial distance $\rho=370\,\mu$m away from the propagation axis. Measurements and force predictions \cite{BareschJASA} are in good agreement. Nevertheless, various improvements could be reached by accounting precisely for the effect of the film on the bead's radial acceleration and measuring the absolute pressure field with better accuracy. 
As in this region the radiation force varies linearly with the radial displacement, this corresponds to an approximate  radial trap stiffness, $\kappa\simeq4$~nN/$\mu\mbox{m}$. The maximum radial displacement reached $\rho_{max}=700\,\mu$m for various particles. If the particle is initially located beyond, the positive radial force will expel it instead. 
This remarkable selectivity feature ensures an accurate operation at the single particle level in media with high densities of particles (see SI Video 2 \cite{Supp}). If a unique particle is trapped at the focus, a potential barrier will keep the others away from the equilibrium position. 
Note that this is not the case for most optical trapping experiments where many particles can quickly collect at the focus for a high concentration of particles \cite{Ashkin86,Neuman}. Nevertheless, if two particles are located  close enough to each other, the beam simultaneously lifts both (see SI Video 3 \cite{Supp}). Systems of two trapped particles are shown in Fig.~\ref{fig:multi}. It can be seen that both particles will react to the primary incident beam and be trapped in a single equilibrium position but can additionally exhibit a mutual interplay. Depending on the relative size of each particle, they can either attract or repulse each other and maintain a separation distance $d\simeq0.25\lambda$ (Fig.~\ref{fig:multi} (III)). While a systematic study of multiple particle interaction was beyond the scope of this study, such systems may offer interesting possibilities to probe the accuracy of recent radiation force calculations for multiple-scatterers \cite{BruusMS}.

To demonstrate transportation, once the particle is picked up, the film is very rapidly moved to another position where the particle is precisely and gently dropped in a cavity (see SI Video 4 \cite{Supp}). The streams generated by the high speed translation of the film ($50$~mm/s) do not affect the position of the trapped particle. Reversely, this observation implies the ability to rapidly move the trapped specimen in the host medium displacing the trapping beam instead and suggests that the tweezers are relevant for high speed manipulation tasks.

Another interesting possibility is to precisely control the dynamics of the particle to fashion an actuator for various applications. A first and simple way to manage this is to control successive cycles of catching and dropping of the particle by switching the power source on and off at different rates. This is illustrated in the see SI Video 5 \cite{Supp}. The incident power is driven with a square signal of three different periods 0.3, 0.15 and 0.1 s. Thus, we can control the amplitude and the frequency of the oscillations of a particle around its equilibrium position. This could be used for instance to initiate streams in the host fluid. While not exploited here, the versatility of the wavefront synthesis technique offers many possibilities to reconfigure in real time the form and the position of the trapping beam as in holographic optical traps \cite{Grier3,Grier,Ruffner2012}.
Note that, in all experiments, once the particle is trapped, we completely remove the film to make sure it does not disturb the trapping force.

  
\paragraph{Discussion}
For the first reported time, the results demonstrate remote trapping of elastic particles in three dimensions with the gradient force of a single acoustic beam, \textit{i.e.} acoustical tweezers. We exerted forces overtaking those of optical tweezers by 3 to 4 orders of magnitude while reducing the intensity flux by 5 orders of magnitude. Acoustical tweezers can selectively pick and control the position of a particle in all three dimensions. As for optical tweezers, a mechanical positioning of the beam focus with a translational stage will ensure a precise control of the particle's position. If the source is made of a set of independent transducers, electronic steering of the beam is feasible such as spatial light modulators in optics \cite{Grier3}. Moreover, acoustical vortices carry a quantified orbital angular momentum \cite{Hefner,MarchianoThomas2003} that can be transferred to matter \cite{Volke,PadgettAcous,Demore,Wunenburger,Marston2011b,Demore2}, so that controlled rotation of absorbing particles is expected to provide a fourth degree of freedom for contactless manipulation and forms the basis of an undergoing study. 

Concerning different materials, the theory predicts that any elastic particle is stably trapped although increasing the acoustic power may be necessary to overcome the gravitational force at this scale. The trapping beam seems robust to other shapes as demonstrated experimentally for a system of two spheres. Regarding the particle size, beads up to a third of the wavelength in diameter were successively trapped but larger ones are predicted to be expelled. Small beads are also trappable. However, the gradient force decreases faster than the acoustic streaming drag force for small particles. The best solution is to scale up the frequency to keep a radius to wavelength ratio sufficiently large. Driving frequency approaching $f=100$~MHz \cite{KKShung} would give access to beads in the microscopic range.

It seems particularly relevant and appealing to extend the ``microscopic handle'' technique to acoustical tweezers to apply and precisely measure forces involved at the cellular scale in biophysical processes as morphogenesis, organogenesis and cell adhesion \cite{Desprat2008,Brunet2013}.

\end{document}